# On the Role of Dissipation in the Early Stages of Relativistic Heavy Ion Collisions

L. Mornas[*], U. Ornik[‡]

*Theory Group, GSI, D-64200 Darmstadt, Germany*
December 6, 1994

## Abstract

The influence of the dissipative terms on the conditions of formation and the characteristic parameters of shock waves in relativistic nuclear collisions is investigated for three types of equation of state (non linear QHD-1, resonance gas and lattice QCD). Energy and velocity profiles are obtained in a one-dimensional model; the duration of the shock phase and width of the shock front are calculated. It is shown that the presence of a phase transition results in a strong enhancement of the width of the shock front, which results in an increase of transparency. This effect, combined with the fact that the nuclei have a finite size, prevents the energy density to rise to its maximum value (full stopping) as would be predicted by a non dissipative shock model.

## 1 Introduction

Among the possible approaches for the description of relativistic heavy ion collisions, hydrodynamics has obtained some success in reproducing experimental data [1] – [6]. It is known from nonrelativistic calculations [7] – [10]

[*]e-mail: mornas@rzri6f.gsi.de
[†]e-mail: ornik@tpri6d.gsi.de
[‡]submitted to Nuclear Physics A

that dissipative effects can have an important influence on the observables. On the other hand, relativistic effects are thought to be essential for a good description of nuclear collisions above 1 GeV energy. However, a fully relativistic hydrodynamic code including dissipative terms yet practically does not exist (see however [11]). The problems hampering this development are of two different natures. First, there is a numerical one: The Lorentz invariance now requires the time and space coordinates to be treated on an equal footing. This introduces time derivatives in the hydrodynamical equations which cannot anymore be solved by the usual explicit algorithms. Secondly, some doubts have been emitted concerning the validity of the Navier-Stokes equations for relativistic problems, as these may exhibit unstable and acausal modes (see *e.g.* [12] – [16]).

In this paper we would like to make a first step in the study of relativistic dissipative hydrodynamics by addressing the question of the structure of relativistic shock waves in the presence of transport terms. This simplified picture of the early stage of the collision allows to find an analytical solution and permits to gain insight into the basic physical processes at hand. The solution of this problem may then be used as a simple picture for the early stage of the collision and as input for the subsequent expansion phase. The model yields energy density and velocity profiles, the duration of the compression phase and an analytical expression for the width of the shock front. If the width of the shock front is larger than the size of the nuclei, the notion of shock looses its meaning. It therefore serves as a criteria to what extent hydrodynamical models are applicable. Moreover, it provides a basic check in the development of a more refined hydrodynamical code and a test of some general features of the equation of state.

## 2 Basic formalism

### 2.1 Relativistic Navier-Stokes equation

The relativistic Navier-Stokes equations [17, 18] are derived from the relation of conservation of the baryon number and energy-momentum

$$\partial_\mu J^\mu = 0 \qquad (1)$$
$$\partial_\nu T^{\mu\nu} = 0. \qquad (2)$$

The baryon current and energy momentum tensor take the general form:

$$J^\mu = \rho u^\mu \tag{3}$$

$$T^{\mu\nu} = \varepsilon u^\mu u^\nu - P\Delta^{\mu\nu} + 2\eta\sigma^{\mu\nu} + \zeta/3\theta\Delta^{\mu\nu} + q^\mu u^\nu + q^\nu u^\mu \tag{4}$$

where $u^\mu = (\gamma, \gamma\vec{v})$ is the hydrodynamical 4-velocity, $\Delta^{\mu\nu} = g^{\mu\nu} - u^\mu u^\nu$, $g^{\mu\nu} = diag(1, -1, -1, -1)$. $\eta$ and $\zeta$ are the shear and bulk viscosities; $q^\mu$ is the heat current and is given by

$$q^\mu = \lambda[\Delta^{\mu\nu}\partial_\nu(T) - T\dot{u}^\mu] = -K\Delta^{\mu\nu}\partial_\nu(\mu/T), \tag{5}$$

$\lambda$ is the thermal conductivity and K is related to it through

$$K = \rho T^2 \lambda/(\varepsilon + P). \tag{6}$$

$\sigma^{\mu\nu}$ is the shear tensor and $\theta$ is the 4-divergence of the velocity;

$$2\sigma^{\mu\nu} = [\Delta^{\mu\alpha}\Delta^{\nu\beta} - \frac{1}{3}\Delta^{\mu\nu}\Delta^{\alpha\beta}](\partial_\alpha u_\beta + \partial_\beta u_\alpha) \tag{7}$$

$$\theta = \partial_\mu u^\mu \tag{8}$$

The thermodynamical properties ($\varepsilon$, $P$, $\rho$) and transport properties ($\lambda$, $\eta$, $\zeta$) of the medium may be given by phenomenological models or derived from an underlying microscopical theory (see below).

## 2.2 Equation of state

In the following, four cases were considered for the equation of state (eos):

In order to have a simple model, we first assumed an equation of the form $P = c^2\varepsilon$ with a constant sound velocity, e.g. $1/3$.

We also took the equation from the Walecka model (here with the exchange of $\omega$ and $\sigma$ mesons only) [27]. We used the non linear version of the QHD-1 model, which gives values for the effective mass and incompressibility coefficients compatible with the experimental values. The parameters

$$m_\sigma = 550\text{MeV}; \quad m_\omega = 783\text{MeV}; \quad g_\sigma = 7.947; \quad g_\omega = 6.706;$$
$$B = -0.17788\ 10^{-1}; \quad C = 0.39674\ 10^{-1} \tag{9}$$

yield [28]

$$m^* = 0.85; \quad K_\infty = 210\text{MeV}, \tag{10}$$

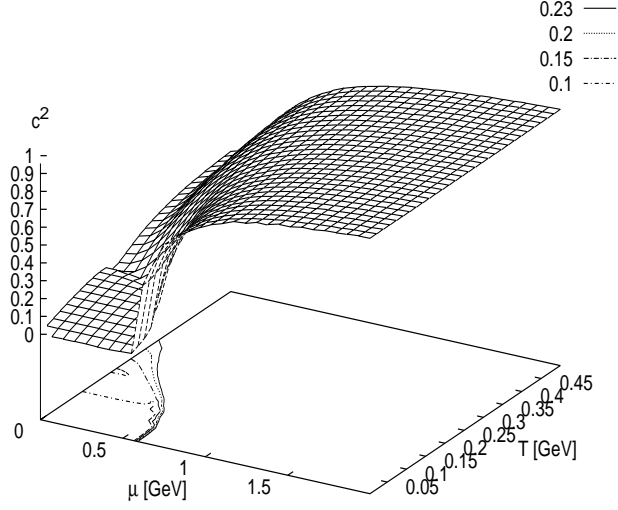

Figure 1: Speed of sound $c^2$ for the non-linear QHD-1 model

in agreement with the current experimental values.

The use of the non-linear QHD-1 model has the advantage that one can derive the relativistic Navier-Stokes equations, the eos and the transport coefficients in a consistent manner (see below).

Next, we used the resonance gas model [29]. In the version calculated here, all particles and resonances of the Particle Data Book are taken to contribute to the pressure and energy density.

$$\varepsilon = \frac{1}{(2\pi^2)} \sum_{i=1}^{200} \int_0^\infty dp\, p^2 \frac{g_i E_i}{\exp\left(\frac{E_i - B_i \mu_b - S_i \mu_s}{T}\right) \pm 1} \qquad (11)$$

$$\rho_b = \frac{1}{(2\pi^2)} \sum_{i=1}^{200} \int_0^\infty dp\, p^2 B_i \frac{g_i}{\exp\left(\frac{E_i - B_i \mu_b - S_i \mu_s}{T}\right) \pm 1} \qquad (12)$$

$$\rho_s = \frac{1}{(2\pi^2)} \sum_{i=1}^{200} \int_0^\infty dp\, p^2 S_i \frac{g_i}{\exp\left(\frac{E_i - B_i \mu_b - S_i \mu_s}{T}\right) \pm 1} = 0 \qquad (13)$$

where $i$ is running over the particle and resonance states, $E_i$ is the energy of particle $i$, $g_i$ is the degeneracy factor, $B_i$ the baryon number and $S_i$ the strangeness, $\mu_b$ and $\mu_s$ are the chemical potentials for baryon number and strangeness respectively, and $\rho_b$ and $\rho_s$ are the corresponding densities.

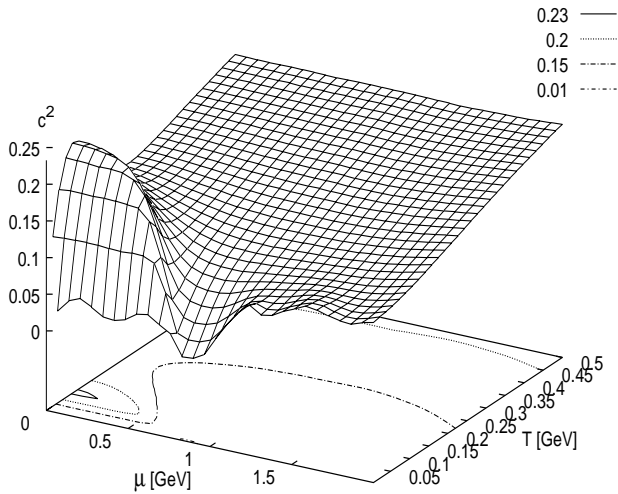

Figure 2: Speed of sound $c^2$ for the bootstrap equation of state

The conservation of strangeness is built in by introducing a strange chemical potential $\mu_s$ which is adjusted in such a way that the strangeness of the system is vanishing. This strange chemical potential grows with baryonic density and temperature (cf Fig 3).

It is interesting to note that, at high baryon densities, $\mu_s$ exceeds the kaon mass. This leads to a condensation of the kaons which we take into account by adding a condensate contribution $n_K^{con}$, which is given by

$$n_K^{con} = -\rho_s(\mu_b, T, \mu_s = m_K)\theta(\mu_s - m_K) \qquad (14)$$

This contribution strongly increases at high temperature and baryonic densities (cf Fig 4)

Finally, we considered an eos derived from lattice QCD calculations including a first order phase transition from a quark-gluon plasma towards

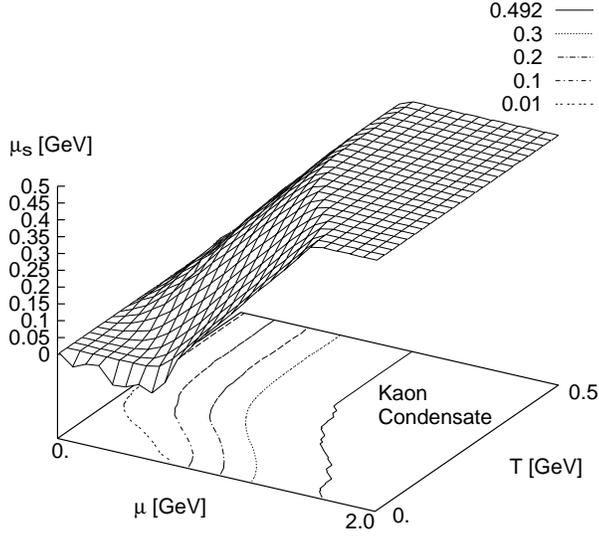

Figure 3: strange chemical potential for bootstrap eos

nuclear matter. The lattice data are reproduced by a phenomenological fitting formula [26]:

$$P = c_0^2(\varepsilon)\varepsilon$$
$$c_0^2(\varepsilon) = \left[\alpha + \beta\tanh\left(\gamma\ln\left(\frac{\varepsilon}{\varepsilon_c}\right) + \delta\right)\right]\left[1 - \frac{(1-\xi)\Gamma^2}{\ln^2(\varepsilon/\varepsilon_c) + \Gamma}\right] \quad (15)$$

Two sets of parameters were used (see Table)

|         | $\alpha$ | $\beta$ | $\gamma$ | $\delta$ | $\xi$ | $\Gamma^2$ | $\varepsilon_c$ |
|---------|----------|---------|----------|----------|-------|------------|-----------------|
| set (1) | 5/21     | 2/21    | 0.24     | 1.05     | 0.3   | 0.73       | 3.0             |
| set (2) | 4/15     | 1/15    | 0.24     | 1.05     | 0.25  | 0.73       | 1.5             |

In Figs. 1, 2, 5, we show the behavior of $c^2 = P/\varepsilon$ as a function of the temperature and of the chemical potential for the various equations of state considered. As we will see later, $c^2$ is the shock velocity in the ultrarelativistic limit.

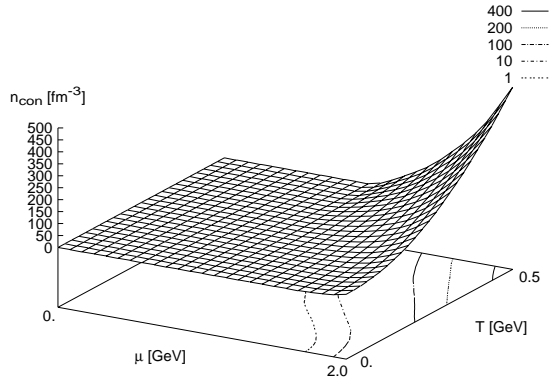

Figure 4: density of kaon condensate for bootstrap eos

## 2.3 Transport properties

The volume viscosity $\zeta$ is in general negligible, and it plays in the 1d case a role similar to that of the shear viscosity (*cf* eqs. (4-7)), so that it will not be considered in the following. We will consider three cases for the thermal conductivity $\lambda$ and the shear viscosity $\eta$.

In order to keep things simple, we first chose constant transport coefficients,

$$\eta = 0.05 \text{GeV}/(\text{fm}^2\text{c}); \quad \lambda = 0.2/(\text{fm}^2\text{c}) \tag{16}$$

Second, in the Walecka model we used a fit of the thermal conductivity and shear viscosity as provided by [31]. These transport coefficients can also be calculated from the Walecka $\sigma - \omega$ model in a consistent manner with eos [33], which gives values somewhat larger but qualitatively similar to those of [31]. In Fig. 6, we show a plot of the viscosity $\eta$, the thermal conductivity $\lambda$ as well as of the parameter $K$ (Eqn. 6) entering the equation for the shock profile as coefficient of $\partial_z(\mu/T)$. These dissipative coefficients are given as a function of the temperature for five values of the baryon density: 0.2, 0.7, 1, 2 and 4 times the saturation density $\rho_0$.

For the lattice eos, we made a fit of the shear viscosity from the results of [31, 34, 35] which reproduces the characteristic $T^3$ behavior above $T_c = 0.2$ GeV [36] and have a sharp increase around $T_c$ [34]. (As a matter of fact, the cross section is expected to present a sharp change at a phase transition, due to the critical opalescence phenomenon. Then, the transport coefficients, which are in first approximation inversely proportional to the cross section, follow this behavior).

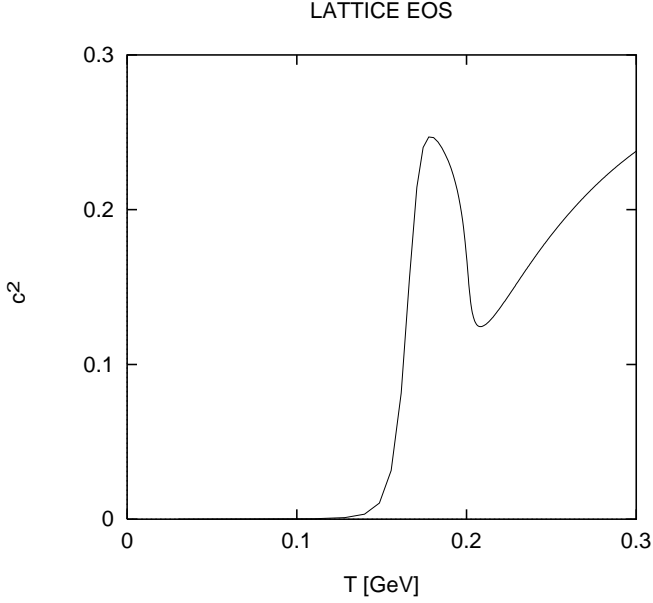

Figure 5: Speed of sound $c^2$ for the lattice eos (set 1)

## 2.4 Shock waves

In the following, we will consider only the one-dimensional problem. In this case, the Navier-Stokes equations take the simple form

$$\partial_t(E) = -\partial_z[(E+P)v + \gamma K \mathcal{D}_q - (4\eta + \zeta)\theta v/3] \qquad (17)$$
$$\partial_t(M) = -\partial_z[Mv + P + \gamma K v \mathcal{D}_q - (4\eta + \zeta)\theta/3] \qquad (18)$$
$$E = (\varepsilon + Pv^2)\gamma^2 + 2K\gamma^3 v \mathcal{D}_q - (4\eta + \zeta)\gamma^2 v^2 \theta/3 \qquad (19)$$
$$M = (\varepsilon + P)\gamma^2 v + K\gamma^3(v^2+1)\mathcal{D}_q - (4\eta + \zeta)\gamma^2 v \theta/3 \qquad (20)$$

with $E = T^{00}$, $M = T^{0z}$, $u^\mu = (\gamma, 0, 0, \gamma v)$, $\theta = \partial_t(\gamma) + \partial_z(\gamma v)$, $\mathcal{D}_q = v\partial_t(\mu/T) + \partial_z(\mu/T)$.

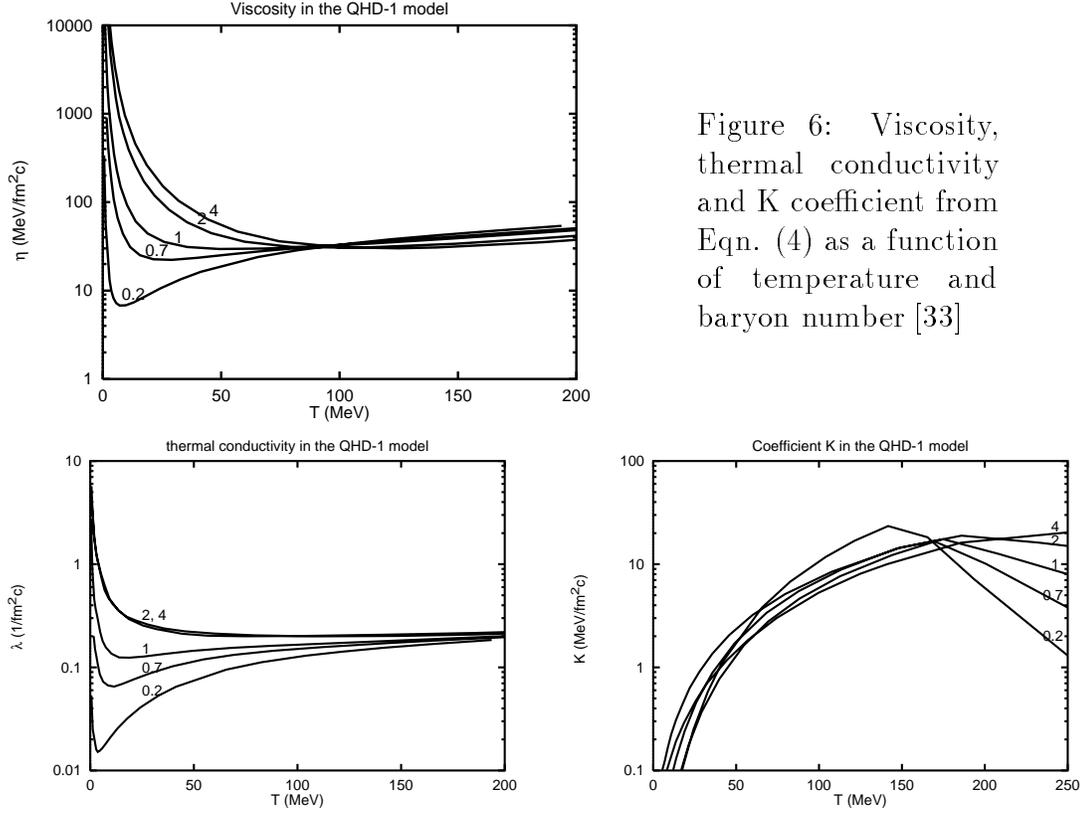

Figure 6: Viscosity, thermal conductivity and K coefficient from Eqn. (4) as a function of temperature and baryon number [33]

We can see that in the simple 1-D case, the viscosity terms formally play the role of an effective pressure

$$P_{eff} = P - (4\eta + \zeta)\theta/3. \tag{21}$$

As $\theta$ remains negative throughout the shock phase, the effective pressure gets higher than the thermodynamical one. We can also define an effective "velocity of sound" $c_{eff}^2 = P_{eff}/\varepsilon$. For too high values of this parameter – which can result from high values of the viscosities or too high values of the gradient –, the solution of the hydrodynamical equations would become problematic and ultimately would go out of the range of validity of the theory. It has been checked in our calculations that $c_{eff}^2$ never exceeds 1 (It stays at the level of a 10% correction). The formal similarity between ideal fluid equations and one-dimensional viscous fluid equations is very probably at the origin of earlier interpretations of the experimental data, which needed in an ideal fluid model harder equations of state (with higher $c^2$) than expected.

In the configuration of a heavy ion collision, a discontinuity in the velocities at the contact surface between two colliding nuclei results in the formation of shock waves which then propagate outwards. ([19] – [21]; [4, 5]).

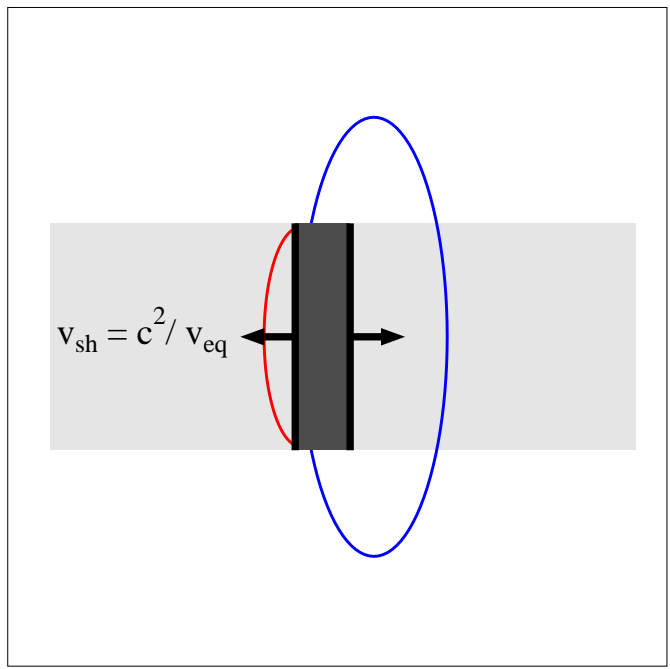

Figure 7: Approximation of the geometry of the nuclear collision through two infinite tubes. The shock front (black bars) is moving with the velocity $v_{sh} = c^2_{max}/v_{eq}$.

In Fig.7 we visualize the basic features of the model. The participant region of the two colliding nuclei is approximated by two semi-infinite tubes of matter. In front of the excited nuclear matter two shock waves are generated, moving apart and propagating into the incoming streams of projectile and target nuclear matter with the velocity $\pm c^2/v_{eq}$. Here, $v_{eq}$ is the velocity of the nuclei in the equal velocity frame.

In the ideal case of two infinitely extended nuclei which we first consider here, a stationary shock establishes. This should remain a good approximation for large nuclei. In this case, the time derivatives vanish in the shock frame, so that we obtain the set of equations

$$\rho \gamma v = \rho_0 \gamma_r v_r \quad (22)$$

$$(\varepsilon + P)\gamma^2 v + \gamma^3 K \mathcal{D}_q (1 + v^2) - \frac{4\eta + \zeta}{3} \gamma^2 v \theta = \text{constant} = \varepsilon_0 \gamma_r^2 v_r \quad (23)$$

$$(\varepsilon v^2 + P)\gamma^2 + 2\gamma^3 v K \mathcal{D}_q - \frac{4\eta + \zeta}{3} \gamma^2 \theta = \text{constant} = \varepsilon_0 \gamma_r^2 v_r^2 \quad (24)$$

which after some rearrangement can be written as a set of differential equations for the velocity and the ratio of chemical potential to temperature

$$\rho \gamma v = \rho_0 \gamma_r v_r \quad (25)$$

$$\partial_z (\mu/T) = [\varepsilon_0 \gamma_r^2 v_r (1 - v v_r) - \varepsilon v]/(\gamma K) \quad (26)$$

$$\partial_z (v) = \frac{\varepsilon_0 \gamma_r^2 v_r [\gamma^2 (1 + v^2)(1 - v v_r) - 1] + \varepsilon v \gamma^2 (c^2 - v^2)}{(4\eta + \zeta)\gamma^5 v/3} \quad (27)$$

Here $\varepsilon_0$ and $\rho_0$ are the energy and baryon number density in cold nuclear matter, $v_r$ is the relative velocity of the inflowing matter and is related to the velocity in the equal velocity frame through

$$v_r = (v_{eq} + v_{sh})/(1 + v_{eq} v_{sh}) \quad (28)$$

$v_{sh}$ being the velocity of the shock.

## 2.5 Asymptotic values

For large enough nuclei, we can consider that the derivatives $\partial_z (v)$ and $\partial_z (\mu/T)$ vanish at $z \to \infty$. The density of the shocked matter reaches an asymptotic value $\varepsilon_{z \to \infty} = \varepsilon_{max}$. Its value its obtained by solving the system of equations (22 – 24) for $v = v_{sh}, \varepsilon = \varepsilon_{max}, \rho = \rho_{max}$ and the derivatives set to zero. In this way we obtain the velocity of the shock

$$v_{sh} = (c_{max}^2 / v_{eq}) \quad (29)$$

with $c^2 = P/\varepsilon$ ($c_{max}^2 = P^{max}/\varepsilon^{max}$). [1] This expression was derived for the case where the pressure in the unshocked matter is vanishing. If this is for some reason not the case, the right hand side of (23) and (24) have to be replaced by $(\varepsilon_0 + P_0)\gamma_r^2 v_r$ and $(\varepsilon_0 + P_0)\gamma_r^2 v_r^2 + P_0$ respectively. Defining $c_{min}^2 = P_0/\varepsilon_0$, the shock velocity is then given as the solution of a $4^{th}$ degree equation obtained by dividing (23) by (24), and replacing $v_r$ by its expression (28). From its solutions, we have to discard three as unphysical. We are left with the fourth solution

$$
\begin{aligned}
v_{sh} &= [c_{max}^2 - c_{min}^2 - v_{eq}^2 + c_{min}^2 c_{max}^2 v_{eq}^2 \\
&+ \sqrt{4(1+c_{min}^2)c_{max}^2 v_{eq}^2 + (c_{min}^2 - c_{max}^2 + v_{eq}^2 - c_{min}^2 c_{max}^2 v_{eq}^2)^2}] \\
&/ [2(1+c_{min}^2)v_{eq}]
\end{aligned} \quad (30)
$$

which is somewhat smaller than in the case $c_{min}^2 = 0$, and tends towards $c^2$ irrespective the value of $c_{min}^2$ when $v_{eq} \to 1$. The expression (30) we derived for the velocity of the shock is more general than the one derived in [18] ($v_{sh} = c_{max}^2$) and extends the model from ultrarelativistic applications to relativistic and semirelativistic problems ( SIS and AGS energies).

---

[1] $c^2$ has to be distinguished from $c_s = \sqrt{\partial P/\partial \varepsilon}|_{s=const}$ (s:entropy-density) which is the speed of sound in the shocked matter. Although our $c^2$ is *not* the speed of sound, we will use this loose way of speaking throughout the paper.

The expression (29) for $v_{sh}$ also provides a lower limit in $E_{lab}$ for the formation of a shock:

$$v_{eq} > c_{max}^2 \Longrightarrow E_{lab} > m(1+c^4)/(1-c^4); \quad (31)$$

e.g. for $c^2 = 1/3$, $E_{lab} > m_n + 0.235\text{GeV}$. Below this energy, we have a simple overlapp of the two densities rather than a shock wave.

Behind the shock, the energy density reaches the asymptotic value

$$\varepsilon_{max} = \varepsilon_0[\gamma_r^2 v_r(1+c_{min}^2)]/[\gamma_{sh}^2 v_{sh}(1+c_{max}^2)] \quad (32)$$

The value $\varepsilon_{max}$ for the energy density behind the shock front is the same, whether there is dissipation or not. This behaviour comes from the assumption of infinitely extended nuclei. In that case, even for a large mean free path (i.e. a large width of the shock front), the instreaming matter would be stopped sooner or later. For more realistic nuclei with finite radii, this is evidently not the case anymore. (c.f. §4)

It is also interesting to express $\varepsilon_{max}$ as a function of $c^2$ and the energy available in the equal velocity system $E_{lab}$, with (for $c_{min}^2 = 0$) $v_{eq} = \sqrt{E_{lab}^2 - m_n^2}/(E_{lab} + m_n)$

$$\varepsilon_{max} = \varepsilon_0 \gamma_{eq}^2 (v_{eq}^2 + c^2)/c^2 \quad (33)$$

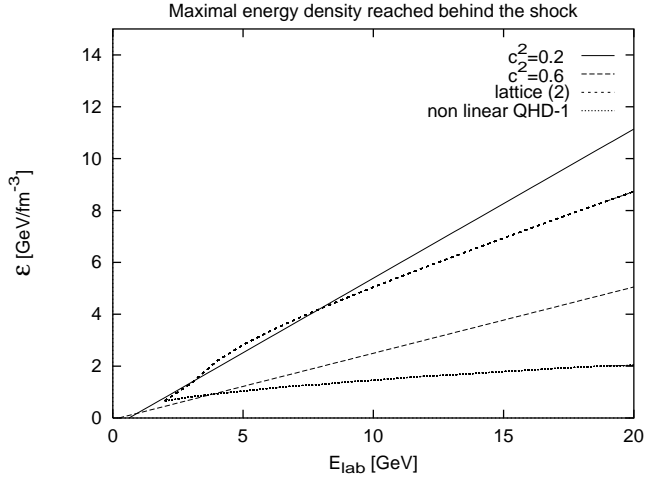

Figure 8: Energy density reached behind the shock as a function of the laboratory energy. The equation of state is here taken to be $P = c_{max}^2 \varepsilon$.

One sees that the maximum reachable energy increases with $E_{lab}$ and will be higher for softer equations of state. (see Fig. 8)

Similarly, one has the asymptotic baryon number density

$$\rho_{max} = \frac{\rho_0 \gamma_r v_r}{\gamma_{sh} v_{sh}} = \rho_0 \gamma_{eq} (v_{eq}^2 + c^2)/c^2 \qquad (34)$$

From this, we also can derive the entropy produced across the shock

$$\Delta S = \left( \varepsilon_{max}(1 + c_{max}^2) - \mu_{max}\rho_{max} \right)/T_{max} \qquad (35)$$

## 3  Applications

### 3.1  Width of the shock front

The effect of the transport terms is to produce a sharp but continuous change of the variables across the shock front instead of a discontinuity.

Let us first neglect the contribution of thermal conductivity. The acceleration of the matter through the shock front is now given by

$$\frac{dv}{dx} = \frac{\varepsilon_0 \gamma_r^2 v_r}{(4\eta + \zeta)/3 \ \gamma^5 v}[(1 + c^2)\gamma^2(1 - vv_r) - 1] \qquad (36)$$

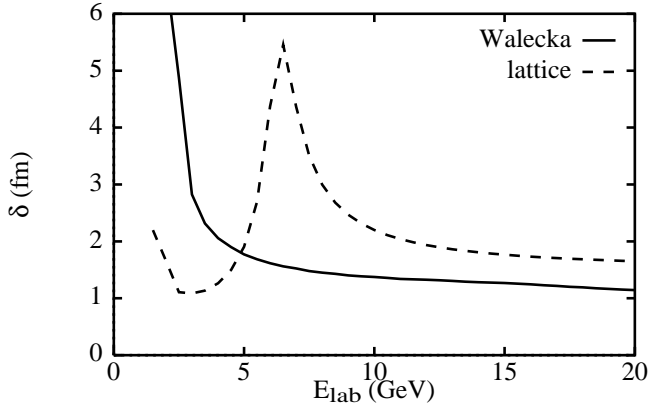

Figure 9: Width of the shock front as a function of the projectile energy

and the energy density given from

$$\varepsilon = \varepsilon_0 \gamma_r^2 v_r (1 - v v_r)/v \tag{37}$$

The characteristic length scale over which the change of the thermodynamical variables takes place – the width of the shock – can be evaluated through

$$\delta = \int_{v_{sh}}^{v_r} (dv/dx)^{-1} dv \tag{38}$$

It is proportional to the viscosity, as can be read from Eqs. (36-38).

The width $\delta$ is plotted in Fig. 9. It is seen that it strongly depends on the eos, i.e. it increases with $c^2$. In general we observe a decrease with the bombarding energy.

The curve for the lattice model with phase transition presents an interesting peculiarity: After falling down to 1 fm, the curve rises again around $E_{lab}$=6.5 AGeV, corresponding to the appearance of the phase transition towards quark gluon matter. The rise is more spectacular when the viscosity also presents a sharp variation at $T_c$, but is still present if the viscosity is supposed to be constant.

This behavior has two distinct origins: First, in the transition region, the density of the matter increases drastically with the collision energy. This corresponds to a quick variation in the parameter $c$, which shape is reflected in the curve for $\delta$. Secondly the width increase at transition is enhanced by dissipation. As a matter of fact, the viscosity is expected to increase strongly in this region, thereby increasing the width of the shock wave.

Later we shall see that this behavior will influence the nuclear transparency in the phase transition region in a characteristic way. (see §4)

## 3.2 Shock profile and the influence of the thermal conductivity

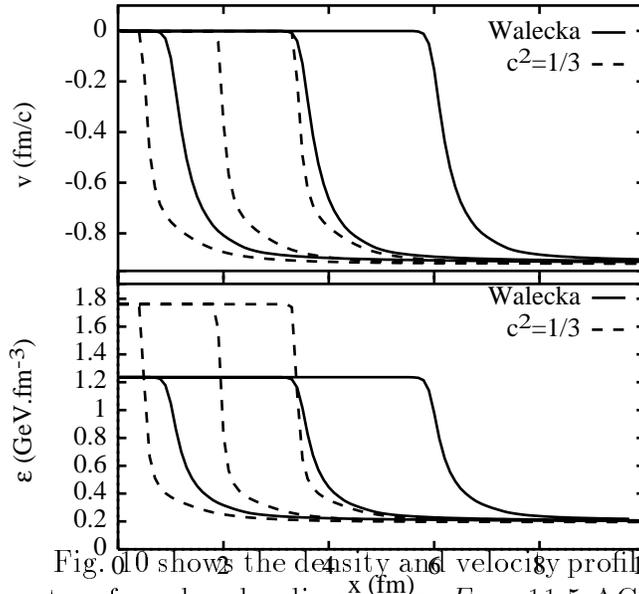

Figure 10: Energy density and velocity profiles in the equal velocity system

Fig. 10 shows the density and velocity profiles as seen in the equal velocity system for a bombarding energy $E_{lab}$=11.5 AGeV (typical at AGS) for three equally distant time steps. The plot was obtained by solving ( 36 – 37), i.e. by neglecting the contribution of the thermal conductivity. We can see that the maximal reachable energy density in the Walecka model is lower than in the case of a constant speed of sound 1/3. Also, the shock wave propagates faster for the QHD-1 eos. As a matter of fact, the average speed of sound in the QHD-1 model is higher than 1/3. Therefore (see 29, 32) $\varepsilon_{max}$ is lower and $v_{sh}$ is higher.

Up to now, we have been neglecting the effects of the thermal conductivity, as is widely done in hydrodynamics calculations with dissipation. We would like to stress here that this assumption might not be justified. The numerical values of the transport coefficients are hardly known experimentally, however the microscopical models [31, 33] rather predict the (shear) viscosity $\eta$ and the parameter K (related to the thermal conductivity, (cf. 6) to be of the same order of magnitude (see Fig. 6 )

Let us study the modification brought about by the taking into consideration of thermal conductivity. In order to do so, we rewrite (26 – 27) in a form similar to (36 – 37) in such a way as to separate the contribution of viscous dissipation, and the additional contribution of the thermal conductivity terms:

$$\varepsilon = \varepsilon_0 \gamma_r^2 v_r (1 - v.v_r)/v - \gamma K \mathcal{D}_q / v \qquad (39)$$

$$\partial_z(v) = \frac{\varepsilon_0 \gamma_r^2 v_r}{(4\eta + \zeta)/3\gamma^5 v}[(1 + c^2)\gamma^2(1 - v.v_r) - 1] + \frac{(v^2 - c^2) K \mathcal{D}_q}{(4\eta + \zeta)/3\gamma^2 v} \qquad (40)$$

Since $\varepsilon$ increases across the shock and $\mu/T$ decreases with increasing $\varepsilon$, the product $K\mathcal{D}_q$ is negative. The additional contribution to $\varepsilon$ is then positive. The additional contribution to $\partial_z(v)$ can be either positive or negative, depending on the sign of $(v^2 - c^2)$. For high velocities it is positive so that we expect the velocity first to drop faster than without thermal conductivity. Then, when $v$ has dropped below $c$, we expect the slope to decrease and $v$ to return slowlier to $v_{sh}$. The global effect is an additional broadening of the shock front and a slight change in the shape of the profiles.

In Fig. 11, we compare the shock profiles obtained with and without taking the thermal conductivity into consideration. The curves were obtained for a laboratory energy $E_{lab} = 4$ GeV, on the right for the non linear QHD-1 eos and on the left for the resonance gas eos. For both calculations, we used the parametrization of the viscosity and thermal conductivity given by Danielewicz [31]. We see that the energy density and velocity profiles have their width increased by say 15 to 20 %. The effect is much more remarkable on the temperature profile. The temperature has already increased much before a notable change has ocurred in the energy density, due to a better transport of $T$ when the thermal conductivity is taken into consideration, whereas changes in $T$ could only be brought about indirectly by the equation of state if only the viscosity was taken.

# 4 Dissipation and transparency

As we already discussed the maximal density (Eq. 32) will only be reached for the collision of infinite tubes of nuclear matter, because the length of the colliding nuclei has to be larger than the width of the shock wave $\delta$. Therefore the maximum density actually reached will depend on the size of the nuclei.

If the width of the shock is larger than the diameter of the colliding nuclei the build up of pressure in the region behind the shock will stop in the moment when the shock is reaching the edge of the projectile or target. This gives a simple explanation for transparency from the viewpoint of dissipative fluid dynamics.

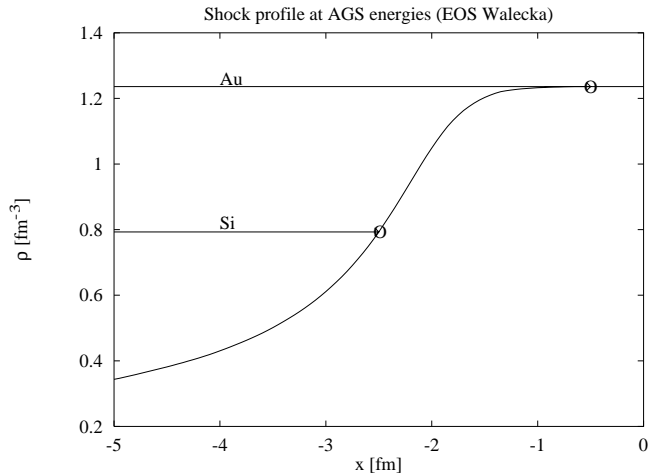

Figure 11: baryon density profile in the case of a nucleus collision at 11.5 AGeV. The horizontal lines are indicating the Lorentz contracted diameters of a Si and a Au projectile

It also gives an estimate of the duration of the shock

$$\tau \sim \frac{(R/\gamma_r)}{v_{eq} + v_{sh}}, \qquad (41)$$

with $R$ being the radius of the colliding nuclei.

In Fig. 12 we compare the density profile of the shock front with the Lorentz contracted diameter of a Si and an Au nucleus. It is seen that in the case of the Si the maximum density is not reachable whereas in the Au case it is possible. One should also note the shape of the shock wave. Over a long distance ($\sim 0.5 fm$) the density increases only very slowly and after that very sharply. This means that the projectile penetrates into the target a remarkable distance before something really happens. As already noticed, in the presence of a first order phase transition, the transparency effect should also lead to observable experimental consequences. The qualitative behavior of the transparency should follow the one of the width $\delta$ (c.f. Fig. 9), *i.e.* we expect the appearance of a sharp increase around the transition temperature. However, due to the uncertainties in the determination of the transition density or temperature, it is not clear where this behavior really starts. Therefore one has to treat our prediction from ( Fig. 9) with care. The position of the bump will shift with the critical density. A closer investigation of transparencies in the range $E_{lab}$ = 5–15 GeV could then give interesting clues as to the equation of state and for what critical temperature and density does the phase transition take place. [2]

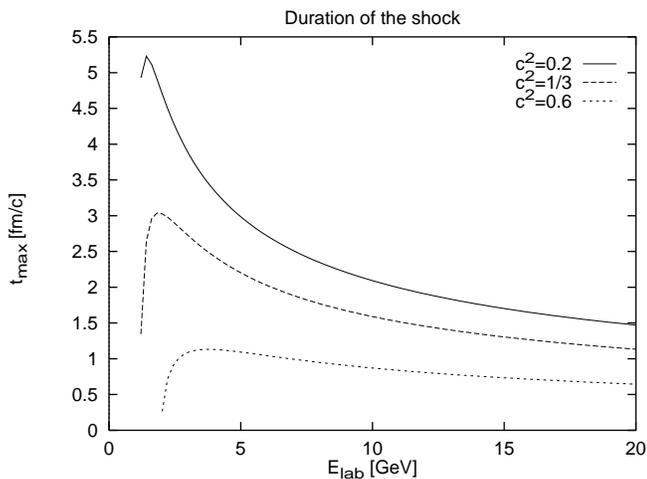

Figure 12: Duration of the collision of two Au nuclei as a function of the laboratory energy

The next two figures (14,15) display the energy density actually reached according to our dissipative shock model in a collision between two nuclei of mass number $A$ in a fixed target experiment at energy $E_{lab}$. This parameter is defined from Fig. 16 as the value at $x = 0$ and $t = t_{max} = (R/\gamma_r)/(v_{eq} + v_{sh})$. We displayed the two cases of a lattice eos and the resonance gas model. In the case of the non-linear QHD-1 model, the energy reaches almost its maximal value eqn. 32 except for a narrow band $A < 10$, due to the irrealistically high value of the velocity of sound of this eos in the high temperature region. For the resonance gas model as well as for the lattice equation of state, we obtain similar shapes. There is a small structure around 3 – 6 GeV laboratory energy, corresponding to the sharp rise and subsequent dip of the speed of sound in this region (see Figs. 2 and 5). The energy then rises slowly with $A$ and $E_{lab}$ for $E_{lab} \sim > 15$ GeV or the lattice eos, the increase is sharper and starting from $E_{lab} \sim > 10$ GeV for the resonance gas eos. As an example, for a Au+Au (resp. Si+Al) collision at AGS energy $E_{lab} = 11.5$ GeV, the energy density reached for our lattice eos is $\sim 1.7$ Gev. (resp. $\sim 0.85$ GeV). Finally we would like to point out that the values given here are probably overestimated, since *(i)* the definition taken here $\varepsilon(x = 0, t = t_{max})$ is the peak value in Fig. 16 which is likely to be rounded off to smaller values in an actual hydrodynamical simulation and *(ii)* the plots displayed in Fig. 14, 15 were made neglecting the thermal conductivity, which will give an additional contribution to the width of the shock front and thus further decrease the actually reachable energy (see Fig. 11).

---

[2]Let us stress again that the phase transition will however not *actually* occur, but rather be postponed to much higher laboratory energies due to the simultaneous increase of transparency

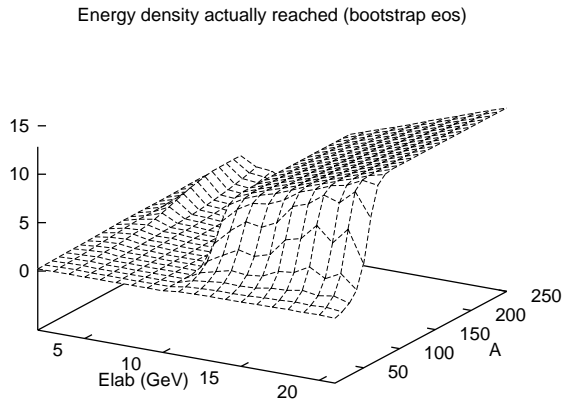

Figure 13: Energy density actually reached in a collision as a function of the mass number of the nuclei $A$ and the laboratory energy $E_{lab}$. The matter is described by the Hagedorn resonance gas model (see §3.1), the viscosity by Danielewicz's formula, $\lambda = 0$

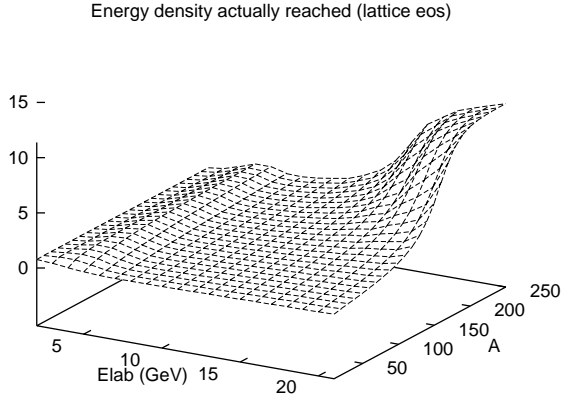

Figure 14: Energy density actually reached in a collision as a function of the mass number of the nuclei $A$ and the laboratory energy $E_lab$. The matter is described by the lattice equation of state (see §3.1), $\eta \propto T^3$, $\lambda = 0$

## 5 Link with the full hydrodynamical simulations

With the help of the parameter $\tau$ defined in (41), we are able to show in Fig. 16 what is to expect for the evolution of the energy density profile when the finite size of the colliding nuclei is taken into account. It is interesting to compare this to the result of a full hydrodynamical calculation (with relativistic ideal fluid hydrodynamics and a numerical viscosity of the order of 0.001 GeV.fm$^{-2}$) [26].

The two models are in quite good agreement which each other; the only noticeable difference is the slope of the energy rise which comes from using different viscosities in the two models: the numerical viscosity used in the hydrodynamical calculation is an order of magnitude lower than that expected in physical situations and used in our shock calculations. The shock model has the advantage of its simplicity, whereas the full hydrodynamical calculations are plagued with numerical problems originating in the size of the gradients involved at this stage. This argues in favour of using the shock model in order to modelize the early stage of a heavy ion collision. The results

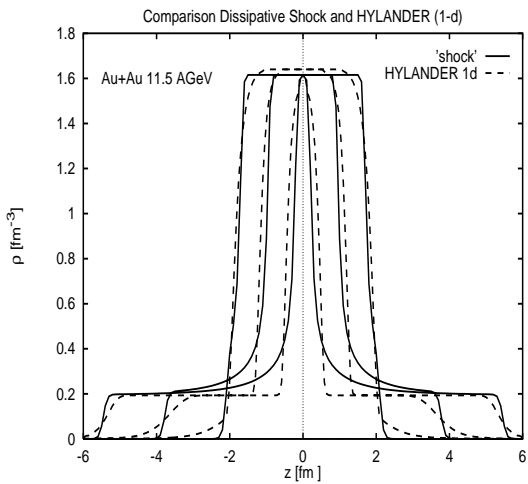

Figure 15: Density profile in the equal velocity system for finite nuclei (here, Au+Au) as given by our shock model (full line) and by a hydrodynamical model (dotted line) [30]

of our shock calculations could then be used as an input for a subsequent 1-D (or possibly 3-D) hydrodynamical calculation in the expansion phase where the steepest gradients are already smoothed out and the equations better behaved.

As an example, we have applied our results as input for a model with Bjorken-like expansion in the 1-D case or in the 3-D case with cylindrical symmetry. The 1-D hydrodynamical equation in the hypothesis of Bjorken scaling $u^\mu = x^\mu/\tau$, $\varepsilon = \varepsilon(\tau)$, $P = P(\tau)$, $\alpha(:= \mu/T) = \alpha(\tau)$ is very simple:

$$\frac{d\varepsilon}{d\tau} + \frac{\varepsilon + P}{\tau} - \frac{4\eta + \zeta}{3}\frac{1}{\tau^2} = 0 \tag{42}$$

Two remarkable properties of this equation should be noticed: (*i*) The thermal conductivity terms disappear from the equation for the 1-D Bjorken model (For a 2-D model with cylindrical symmetry and longitudinal Bjorken scaling, this is not the case any more) (*ii*) Terms containing derivatives of the viscosities with $\tau$ also disappear.

If the equation of state can be put into the form $P = c_s^2 \varepsilon$ and $c_s$, $\eta$, $\zeta$ are assumed to be independant of $\tau$, (??) admits an analytical solution

$$\varepsilon = \left(\varepsilon(\tau_0) - \frac{4\eta + \zeta}{3}\frac{1}{c_s^2\tau_0}\right)\left(\frac{\tau}{\tau_0}\right)^{-(1+c_s^2)} + \frac{4\eta + \zeta}{3c_s^2\tau} \tag{43}$$

In this equation, $\tau_0$ represents the instant at which the expansion phase starts and has to be equated with the duration of the shock phase (41).

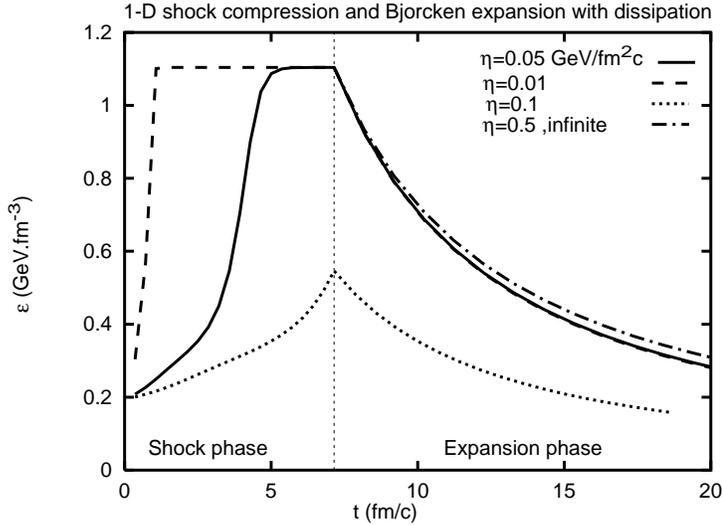

Figure 16: 1-D Bjorken expansion with dissipation

$\varepsilon(\tau_0)$ is the maximal energy reached in the collision (see Figs. 14-15). In Fig. 17, we display numerical results for the evolution of the energy density for Au+Au at $E_{lab} = 4$ GeV through the compression (shock) phase and subsequent Bjorken-expansion as given by (43). Four cases were considered: a small viscosity $\eta = 0.01$ GeV/fm$^2$c (full line), a viscosity in the order of magnitude of what is predicted by microscopical models $\eta = 0.05$ GeV/fm$^2$c (dashed line), a high viscosity $\eta = 0.1$ GeV/fm$^2$c (dotted line). These three cases show an important difference in the shock phase. Indeed, the finite width of the shock front arising from viscosity strongly modifies the rate at which the energy density is rising. On the contrary, in the Bjorken-expansion phase the rate of decrease of the energy density depends very weakly of the value of the viscosity, so that the continuous and dashed lines are practically undistinguishable on the figure. In order to make a comparison, we also show (dot-dashed line) the Bjorken expansion with an unrealistically high value of the viscosity $\eta = 0.5$ GeV.fm$^2$c, starting from the maximal energy density (i.e., for infinitely extended nuclei in this last case). The equation of state is assumed to be given by $P = \varepsilon/3$. From this very simplified model, one would thus conclude that the influence of the dissipation is not very important in the expansion phase. However, the Bjorken scaling assumption is probably a too rough hypothesis.

# 6  Conclusion and outlook

Let us first shortly summarize the results presented in this paper.

We solved the 1-D relativistic dissipative hydrodynamical equations for a stationary shock. When only viscous dissipation is present, a shock width could be defined by an analytical expression. The width is decreasing rapidly with increasing laboratory energy; it is proportional to the viscosity and strongly dependent on the stiffness of the equation of state. It was shown to present a strong enhancement in the range 5 –12 GeV, where a phase transition to QGP would occur *if* the colliding nuclei were of infinite extension, so that the matter behind the shock front had enough time to reach $\varepsilon_{max}$. However, the finite size of the nuclei prevents this maximal energy to be reached, instead we expect them to pass partly through each other. This could be an explanation for an unexpectedly high transparency observed at AGS energy (11.5 GeV) a few years ago [37].[3]

The calculations were also made in case of a non vanishing thermal conductivity. It was found that the width of the energy density and velocity profiles are slightly increased. The temperature profile is much broader. The general conclusions as to the enhancement of the shock width and transparency are qualitatively not modified.

Some questions which were not mentioned in the main text are worth being treated here. First, there is the problem of the stability of the shocks. Several non dissipative calculations ([22] – [25]) predict the shock to develop an instability and split instead into two shock fronts. In order to see if this is the case in our lattice eos, we display in Fig. 18 the Taub adiabat. The Taub adiabat is given by plotting the pressure P behind the shock as a function of a parameter $x$. When the system has a non vanishing baryon number $\rho$, $x$ is commonly defined as

$$x = (\varepsilon + P)/\rho^2 \qquad (44)$$

The equation of the Taub adiabat is obtained by eliminating the velocities from eqns. (22 – 24) with the gradients set to zero. The equation of the

---

[3]According to Fig. 12, we would estimate the transparency to be $K \sim \rho(Si)/\rho(Au) = 0.66$. This is comparable with the result of [37] who find that 11 out of 28 nucleons passed through the target without being affected by the collision, corresponding to $K \sim 0.61$. We do not think however that this coincidence should be taken too seriously.

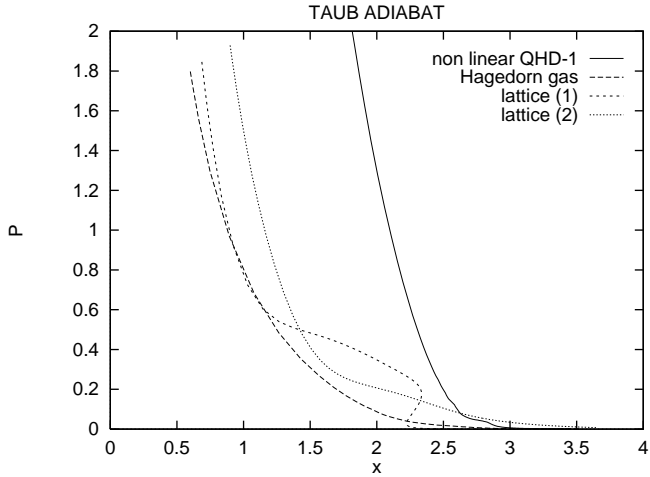

Figure 17: Taub adiabat for various choices of the equation of state

Taub adiabat is then found to be (the index 1 (2) describes the unshocked (shocked) matter respectively)

$$(\varepsilon_2 + P_2)x_2 - (\varepsilon_1 + P_1)x_1 = (P_2 - P_1)(x_2 + x_1) \tag{45}$$

The quantities before the shock (index 1) and the equation of state behind the shock being known, (45) is an implicit equation for $P_2(x_2)$.

However, when the baryon number is vanishing such as is the case for our lattice eos, $x$ has to be defined in an other way. It was proposed by Danielewicz and Ruuskanen [32] to use instead

$$x = [(\varepsilon_2 + P_2)\gamma_2^2 v_2^2]/[(\varepsilon_1 + P_1)\gamma_1^2 v_1^2] \tag{46}$$

One obtains in this case formally the same equation as in (45) with the replacement $x_1 \to 1$.

When moreover the pressure in the unshocked region vanishes, as is the case for cold matter at nuclear density (and as was assumed throughout this paper), the Taub adiabat formula simplifies to a transparent form

$$P_{max} = \frac{\varepsilon_0}{[\frac{(x/x_0)}{c_{max}^2} - 1]} \tag{47}$$

with $x_0 = \varepsilon_0/\rho_0^2$ for nonvanishing baryon number, 1 otherwise.

In order for the system to be stable against the formation of a double shock, the pressure has to be a monotonously decreasing function of $x$ between $x_{max} = x_0 \, (v_s/v_r) = x_0 \, (c^2/v_{eq}^2) \, (v_{eq}^2 + c^2)/(1 + c^2)$ and $x_0$. One sees that this is the case for the non-linear QHD, the Hagedorn resonance gas model and the lattice model for parameter set 2. For the lattice eos with parameter set 1, the pressure is not a monotonous function of $x$ anymore , so that a double shock could form in principle. However, this is not so simple in the case of a dissipative shock because it smears out the sharp separation in two shock fronts seen in the case of an ideal fluid [22]–[25] , so that what we observe is a single shock front with an enhanced width rather than a double shock.

We would also like to say a word about the principle question raised by *e.g.* Olson and Hiscock [16], (see also [12] – [15]): It has been argued that the relativistic Navier-Stokes equations in their usual form (eqs. 5–8) suffer from acausality and instability. ("acausal propagation of heat"). We are conscious of this difficulty. However, we think that, for practical purposes, the relativistic Navier-Stokes equation still has a long life expectancy. Of the alternative theories proposed, the method of moments with inclusion of higher order derivative terms in the right hand side of (eqs. 3,5,6) *à la* Israel and Stewart [13] or non local terms [14], none of them is yet 100% satisfactory, and would need further investigation. They have the big disadvantage of introducing additional complexity through the higher order derivatives, while even numerical hydrodynamics using the simpler Navier-Stokes equations are still in their infancy. 9 additional parameters have to be introduced in the Israel theory as coefficients of the higher order terms, which values are not or very bad known theoretically nor measurable experimentally in a foreseeable future. The uncertainties on these parameters again would introduce uncertainty in the theory.

There are other consequences of the finite size of the nuclei which we did not mention in the main part of this paper. The two most important ones are *(i)* The evaporation of particles in a preequilibrium stage through the surface [38] *(ii)* The reduction of the transport coefficients from the finite size effects (see *e.g.* [39]). As a matter of fact, the transport coefficients displayed in Fig. 6 were calculated in a model of infinite nuclear matter. In finite matter, the particles situated near the surface of the nuclei have their mean free path limited in one direction and collide with the potential wall of the nucleus rather than with other particles. This results is smaller values of the dissipative coefficients (as measured for example from fission necks or "hot spots") near the surface, *i.e.* (see *e.g.* Fig. 11) in a region of smaller temperature, thus counteracting the rise at low $T$ ($\eta \propto 1/T^2$) exhibited by the kinetic solution of [31, 33] (see Fig. 6).

Finally, we would like to mention other important consequence of the model concerning the emission of Bremsstrahlung from the acceleration through the shock. From the characteristics of the Bremsstrahlung spectrum, one can have access to the energy density reached, the width of the shock front, and the duration of the shock phase. This in turn may give information about the stiffness of the equation of state and dynamical properties of the matter. This question is currently under investigation and will be the subject of a forthcoming paper.

# 7  Acknowledgements

This work was supported by the Gesellschaft für Schwerionenforschung.

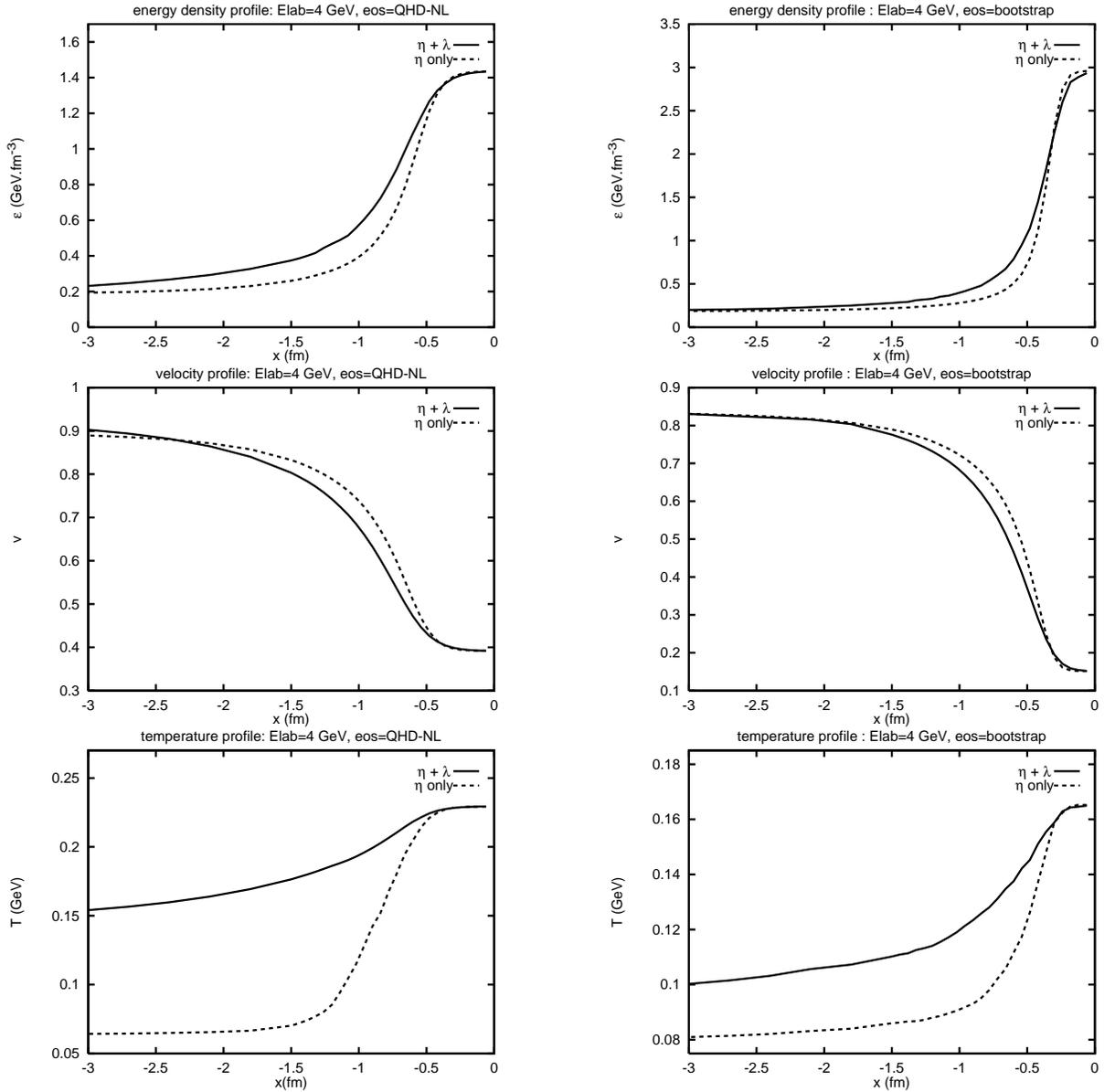

Figure 18: Energy density, velocity and temperature profiles in the rest frame of the shock, for the non linear QHD-1 and the resonance gas equations of state, at $E_{lab} = 4$ GeV. The profiles obtained by taking the both the viscosity and thermal conductivity into account (full line) are compared with those obtained by considering the viscosity only (dashed line)